\documentclass[runningheads,orivec]{llncs}

\usepackage[T1]{fontenc}
\usepackage{graphicx}
\usepackage{hyperref}
\usepackage{color}

\usepackage{bookmark}
\hypersetup{
  pdftitle={Domain-Informed Representation for Evolutionary Sieving in Integral and Module Lattices},
  pdfauthor={Ahmad Tashfeen and Qi Cheng},
  pdfcreator={Ahmad Tashfeen and Qi Cheng},
  pdfproducer={pdfTeX 3.141592653-2.6-1.40.24 (TeX Live 2022)}
  pdfkeywords={Shortest Vector Problem, Integral Lattice, Module Lattice, Sieving, Genetic Algorithm, Post Quantum Cryptography},
  colorlinks,
  linkcolor={blue},
  citecolor={blue},
  urlcolor={blue}
}
\usepackage[plain]{algorithm}
\usepackage{enumitem}
\usepackage[caption=false]{subfig}
\usepackage{amssymb}
\usepackage{mathtools}
\usepackage{booktabs}
\usepackage[
    top=0.8in,
    bottom=0.8in,
    left=0.65in,
    right=0.65in,
    paperwidth=6.11in,
    paperheight=9.26in
]{geometry}

\let\oldparagraph=\paragraph
\renewcommand\paragraph[1]{\oldparagraph{#1.}}

\DeclarePairedDelimiterX{\norm}[1]{\lVert}{\rVert}{#1}
\graphicspath{{plot/last_draft/}{plot/icon/}}

\DeclareRobustCommand{\DO}{\mathbf{do\;}}
\DeclareRobustCommand{\IF}{\mathbf{if\;}}
\DeclareRobustCommand{\THEN}{\mathbf{then\;}}
\DeclareRobustCommand{\ELSE}{\mathbf{else\;}}
\DeclareRobustCommand{\WHILE}{\mathbf{while\;}}
\DeclareRobustCommand{\FOR}{\mathbf{for\;}}
\DeclareRobustCommand{\YIELD}{\mathbf{yield\;}}
\DeclareRobustCommand{\GM}{\mathbf{GramSchmidt}}
\DeclareRobustCommand{\SL}{\mathbf{Select}}
\DeclareRobustCommand{\SWAP}{\mathbf{Swap}}
\DeclareRobustCommand{\FIT}{\mathbf{Fitness}}
\DeclareRobustCommand{\CROSS}{\mathbf{Cross}}
\DeclareRobustCommand{\MUTATE}{\mathbf{Mutate}}
\DeclareRobustCommand{\Elite}{\mathbf{Elite}}

\DeclareRobustCommand{\G}{\Z[i]}
\DeclareRobustCommand{\L}{\mathcal{L}}
\DeclareRobustCommand{\B}{\mathcal{B}}
\DeclareRobustCommand{\LLL}{\text{LLL}}
\newcommand{\NP}{$\mathcal{NP}$}
\renewcommand{\P}{$\mathcal{P}$}
\newcommand{\SVP}{\textsf{SVP}}
\newcommand{\CVP}{\textsf{CVP}}
\newcommand{\SBP}{\textsf{SBP}}
\newcommand{\appr}{\textsf{appr}}
\DeclareMathOperator{\proj}{proj}
\DeclareMathOperator{\spn}{span}

\ExplSyntaxOn
\NewDocumentCommand\col{m} {
  \begin{bmatrix}
    \clist_use:nn {#1} { \\ }
  \end{bmatrix}
}
\ExplSyntaxOff

\DeclareRobustCommand{\bb}[1]{\mathbb{#1}}
\DeclareRobustCommand{\cal}[1]{\mathcal{#1}}
\DeclareRobustCommand{\sc}[1]{\textsc{#1}}

\newcommand*{\ie}{\leavevmode\unskip, {i.e.,} \ignorespaces}
\newcommand{\nil}{\varnothing}
\AtBeginDocument{\def\O{\cal{O}}}
\AtBeginDocument{\def\C{\bb{C}}}
\newcommand{\R}{\bb{R}}
\newcommand{\Z}{\bb{Z}}
\newcommand{\N}{\bb{N}}
\newcommand{\ra}{\rightarrow}

\newcommand{\ep}{\varepsilon}
\renewcommand{\phi}{\varphi}

\newcommand{\near}[1]{\left\lfloor#1\right\rceil}
\newcommand{\arr}[1]{\left\langle#1\right\rangle}
\newcommand{\paren}[1]{\left(#1\right)}

\newcommand{\abs}[1]{\left|#1\right|}
\newcommand{\curl}[1]{\left\{#1\right\}}

\begin{document}

\title{\texorpdfstring{%
      Domain-Informed Representation \\
      for Evolutionary Sieving in Integral \\
      and Module Lattices%
  }{%
      Domain-Informed Representation for Evolutionary Sieving in Integral and Module Lattices%
  }
}
\titlerunning{Evolutionary Lattice Sieving}
%
\author{%
  Ahmad Tashfeen\textsuperscript{(\includegraphics[height=0.6em]{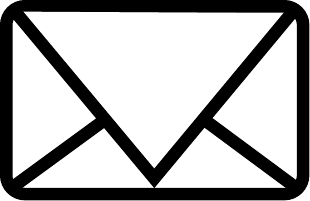})}%
  \href{https://orcid.org/0009-0004-4301-6923}{\includegraphics[height=0.3cm]{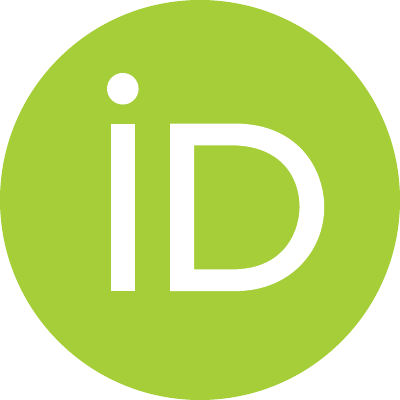}} \and%
  Qi Cheng%
  \,\href{https://orcid.org/0000-0003-4336-3082}{\includegraphics[height=0.3cm]{orcid}}
}

\authorrunning{A. Tashfeen and Q. Cheng}

%
\institute{University of Oklahoma, Norman OK 73019, USA \\
\email{\{tashfeen,qcheng\}@ou.edu}}
\maketitle              
\begin{abstract}
  Traditional cryptography, rooted in problems, e.g., integer
  factorisation or discrete log, is inevitably vulnerable to a fully
  operational quantum computer. Although it remains an engineering
  frontier, the looming threat extends to encrypted data stored today,
  which could be decrypted in the future with quantum capabilities. To
  safeguard against this eventuality, the backbone of the modern
  quantum-safe cryptography is the \emph{Shortest Vector Problem}
  (\SVP{}).  We enhance Laarhoven's treatment of Ajtai et al.'s
  \emph{sieving} as a genetic algorithm (GA) for the \SVP{} by
  incorporating domain-informed \SVP{} representation and crossover
  while naturally extending application to the module lattices.

\end{abstract}
%
%
%


\section{Introduction}\label{intro}%
Lattices model various real-world systems, e.g., relay circuits,
plumbing lines, public roads and digital networks
\cite{Lidl1984,abramkina2024implementation}. Recently, they have gained
significance in post-quantum cryptography, as traditional encryption
methods like RSA are susceptible to quantum attacks. Although fully
operational quantum computers may be distant, existing data could be
compromised in the future. In 2016, 
NIST\footnote{U. S. Department of Commerce’s National Institute of
  Standards and Technology} launched an initiative
\cite{federalRegister2016} to select and standardise quantum-resistant algorithms.
Three out of four chosen final algorithms are based on problems from the
lattice theory
\cite{alagic2019status,alagic2020status,alagic2022status,alagic2025status}:
\mbox{CRYSTALS--K\sc{yber}} and Saber use module lattices as a source
of their hardness and NTRU is based on the Shortest Vector Problem
(\SVP{}). Along with the Knapsack Problem's reduction to the \SVP{},
the \SVP{} has attracted the attention of many prominent
mathematicians over time including Gauss. Today, the most practical
\cite{cryptoeprint:2025/304,zhao2025sieving} approaches to the \SVP{}
are \emph{sieving} and \emph{lattice reduction} algorithms
(Section~\ref{sec:lll}). These include the Lenstra, Lenstra, and
Lov\'asz (\LLL{}), or the block Korkin--Zolotarev (BKZ) algorithms.

More than half of the solutions for the Darmstadt's \SVP{} challenges,
posted at their website \cite{LatticeChallenge2025} are done via some
sieving technique \cite{sun2020review}. Laarhoven's
\cite{laarhoven2019} treatment of sieving as a genetic algorithm (GA)
enabled the application of the evolutionary framework to lattices. In
this work, we enhance and further test this framework by incorporating
field-specific knowledge, while also extending it to the increasingly
significant module lattices.

The primary aim of this paper is to focus on demonstrating the
improvements achieved in a \emph{simple genetic algorithm} only by
integrating insights from lattice theory\footnote{Optimisations
  through the latest machine learning techniques remain an excellent
  opportunity for future research.}. This choice allows our results to
be directly comparable to Laarhoven's \cite{laarhoven2019} simple
genetic algorithm and lets us attribute any improvements solely to the
domain-informed problem representation and crossover.

Our \emph{contributions} are summarised as follows,
\begin{enumerate}
  \item We define an improved \SVP{} representation for GAs enabling a
        versatile crossover operator using the insights from traditional
        algorithms like \LLL{}. Our crossover is compatible with module
        lattices therefore extending a sieving algorithm for module
        lattices for the first time.
  \item We solve all the challenge \cite{LatticeChallenge2025} and
        randomly generated problems till 100 dimensions improving
        application \& scalability of Laarhoven's approach
        \cite{laarhoven2019}.
\end{enumerate}

\paragraph{Outline}
Sec.~\ref{intro} and \ref{prelim} present the current \SVP{}
relevance, key lattice intuitions, heuristics, and notations utilized
in this paper. Sec.~\ref{rw} reviews related
work. Sec.~\ref{sec:method} presents our primary
GA. Sec.~\ref{results} details and discusses the experiments
and their outcomes. Finally, Sec.~\ref{sec:conclusion} provides ideas
for future-work and concluding remarks.


\section{Preliminaries}\label{prelim}%
\begin{figure}
  \begin{minipage}{.5\textwidth}
    \centering
    \includegraphics[width=0.9\linewidth]{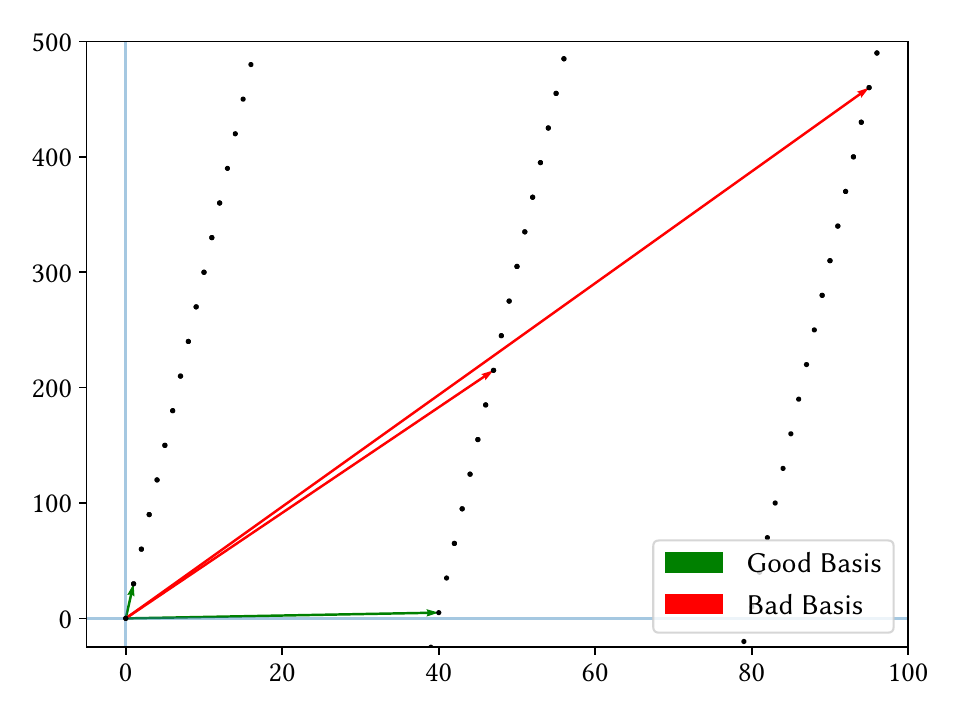}
    \caption{Example of a 2D lattice.}
    \label{2dlattice}
  \end{minipage}%
  \begin{minipage}{.5\textwidth}
    \begin{equation}\label{running}
      \begin{aligned}
        \B_\text{bad}  & = \begin{bmatrix} 95  & 47  \\ 460 & 215 \end{bmatrix} \\
        \B_\text{good} & = \begin{bmatrix} 1  & 40 \\ 30 & 5 \end{bmatrix}
      \end{aligned}
    \end{equation}
  \end{minipage}
\end{figure}

A \emph{lattice} is a discrete additive subgroup of a $d$-dimensional
Euclidean space, i.e., a subset closed under addition where each
element has no other element within a certain radius.
In two dimensions, a lattice maybe colloquially referred to as a grid
of points. See Figure~\ref{2dlattice} for example. Let $\N, \Z, \R,$
and $\C$ be the natural, integer, real and complex numbers
respectively. Consider a basis matrix $\B \in \R^{d \times d}$ with
$d$ linearly independent column vectors
$\{b_i \in \R^d : 1 \leq i \leq d\}$ such that $b_i$ is the
$i^\text{th}$ column of $\B$. We recall a possibly transformed $d$
dimensional Euclidean space as $\{\B x : x \in \R^d\}$, if we restrict
$\B x$ to only the integral linear combinations $x\in \Z^d$, we obtain
a lattice $\L(\B) = \{\B x : x \in \Z^d\}$.  The \emph{dimension} of a
lattice is $\dim(\L) = d$ \ie the number of column-vectors in the
basis matrix $\B$ and the determinant similarly
${\det \L(\B) = \abs{\det \B}}$. If $\B$ also has $d$ number of rows,
we call $\L$ a \emph{full rank lattice}. $\R^d$ is now the
\emph{ambient space} for the reduced set of vectors that are in the
lattice $\L$. The \emph{group operation} of a lattice is the vector
sum, $v + u \in \L$ such that $u,v \in \L$.

\paragraph{Integral Lattices}
If we further restrict $\B \in \Z^{d \times d}$, the resulting lattice
is known as an \emph{integral lattice}. You might imagine a trivial
lattice $\Z^2$ spanned by the identity matrix. We show the plot of a
two dimensional lattice ($d=2$) in Figure~\ref{2dlattice} spanned by
both of the basis stated in Equation~\ref{running}.

\paragraph{Good vs. Bad Basis}
Figure~\ref{2dlattice} shows a ``good'' and a ``bad'' basis given in
Eq.~\ref{running} that span the same lattice. A basis gets better as
its vectors get shorter and more orthogonal and vice versa. The
process of turning a bad basis into a good basis is also referred to
as performing \emph{lattice reduction}.

\paragraph{Module Lattices}\label{sec:module}
Here we define the \emph{module} lattices
\cite{cryptoeprint:2025/304} formed over the \emph{Gaussian integers}
$\G = \{\alpha + \beta i : (\alpha, \beta)\in \Z^2\}$ where
$i^2 = -1$. Despite the name, Gaussian integers are complex numbers
$\alpha + \beta i$ where both, the real
$\Re(\alpha + \beta i) = \alpha$ and the imaginary
$\Im(\alpha + \beta i) = \beta$ parts are limited to the integers. We
can similarly form a lattice, this time in the ambient space of
complex numbers $\C$, by letting $\B \in \G^{d \times d}$ for the
lattice ${\L = \{ \B x : x \in \G^d\}}$.  The norm of a $v \in \G^d$
is given as
$ \norm{v}_2^2 = v^Hv = |v_1|^2 + |v_2|^2 + \cdots + |v_d|^2 $ where
$|v_j|^2 = \alpha^2 + \beta^2$ for $v_j = \alpha + \beta i$. The
vector $v^H$ is known as the \emph{conjugate transpose} of $v$.

\paragraph{Hadamard Ratio}
The
$\mathcal{H}(\B) = \sqrt[d]{\det
  \L(\B)/\norm{b_1}\norm{b_2}\cdots \norm{b_d} }$ is used as
a measure of orthogonality in-between the basis vectors $b_i \in
\B$. Put simply, the closer this ratio is to 1 for a basis $\B$, the
more orthogonal it is.

\subsection{Lattice Problems}

Both the \SVP{} and the Closest Vector Problem (\CVP{}) are
\NP-hard\footnote{However, \SVP{} is \NP-hard under a randomised
  reduction.}. Let $\norm{v}$ be the $\ell^2$ norm of a vector $v$.

\begin{definition}[Shortest Vector Problem (\SVP)]
  Find the shortest non-zero vector $v \in \L$ such that for all
  $u \in \L$ where $v \neq u$, we have $\norm{v} \leq \norm{u}$.
\end{definition}

\begin{definition}[Approximate Shortest Vector Problem (\appr\SVP)]\label{def:alpha}
  Let $\alpha \geq 1$ be an approximation factor and find a $v \in \L$
  such that $\norm{v}$ is no bigger than $\alpha$ times the length of
  the shortest non-zero vector \ie
  $\norm{v} \leq \alpha\norm{v_\text{shortest}}$.
\end{definition}

\begin{definition}[Closest Vector Problem (\CVP)]
  For $w \in \R^d$ where $w \not\in \L$ find a vector $v \in \L$ that
  is closest to $w$ and $\norm{w - v} > 0$ \ie
  $\min_{v \in \L} \norm{w - v}$.
\end{definition}

\begin{definition}[Shortest Basis Problem (\SBP)]
  Find a $b_i \in \B_\text{good}$ from $\B_\text{bad}$ where $b_i$ are
  short in some sense, e.g., minimise
  $\max_{1\leq i\leq d}\norm{b_i}$ or $\sum_{i=1}^d\norm{b_i}^2$.
\end{definition}

\SBP{} may be solved via various lattice reduction algorithms (Section
\ref{sec:reduction}). Any solution to the \SBP{} often also uncovers a
comparable solution to the \appr\SVP. Essentially all lattice-based
cryptography (as talked about in the introduction) relies on the
inability of lattice reduction algorithms to solve the \appr\SVP{}
with an approximation of $\alpha = \O(\sqrt{d})$
\cite{hoffstein2008introduction}.

\subsection{Gaussian Heuristic for the Shortest Vector}\label{sec:gh} It is
difficult to verify a solution for the \appr\SVP{} and \SVP{} since
$\norm{v_\text{shortest}}$ is unknown in the general case. As the
dimension $d$ of a random lattice gets sufficiently large, we may rely
on the Gaussian expected shortest length, i.e., the Gaussian heuristic
where
$\norm{v_\text{shortest}} \approx \sigma(\L) = \sqrt{{d}/{2\pi
    e}}(\det \L)^\frac{1}{d}$ \cite{hoffstein2008introduction}.

\subsection{Genetic Algorithms}\label{sec:gen-ga}

A \emph{genotype} encodes a solution. Genotype instances are called
\emph{individuals}. A set of individuals is a \emph{population}
$u, v \in P$. Populations of individuals evolve through
\emph{generations} via reproduction. Using a \emph{selection}
$\SL(P, \FIT)$ process based on a \emph{fitness-function}
$\FIT: P \ra [0, 1]$, two parent individuals $u, v$ are selected for
\emph{crossover} to produce a child individual
$t \leftarrow \CROSS(u, v)$ in the next generation $R$. The child $t$
may $\MUTATE(t)$. See Alg.~\ref{ga} from Russell
\cite{russell2016artificial}.

\begin{algorithm}
  \begin{flushleft}
    \noindent\textbf{Input}:
      Initial population: $P$, fitness-function: $\FIT: P \ra [0, 1]$. \\
    \noindent\textbf{Output}:
      The best individual in population, according to the fitness.
  \end{flushleft}%
  \begin{enumerate}[label=\arabic*:,noitemsep,partopsep=0pt,topsep=0pt,parsep=0pt]%
    \item $\DO$
    \item $\quad R \leftarrow \nil$
    \item $\quad \FOR (u, v) \leftarrow \SL(P, \FIT)$
    \item $\quad\quad t \leftarrow \CROSS(u, v)$
    \item $\quad\quad \IF (\text{small random probability})\;\THEN t \leftarrow \MUTATE(t)$
    \item $\quad\quad R \leftarrow R \cup \{t\}$
    \item $\quad P \leftarrow R$
    \item $\WHILE (\forall v \in P, \FIT(v) < 1 - \ep)$
  \end{enumerate}%
  \caption{%
    This simple base GA is the same as Laarhoven's base GA \cite{laarhoven2019}
    therefore, attributing all improvements in comparison solely
    to our domain-informed problem representation and crossover operator.
  }%
  \label{ga}%
\end{algorithm}%

 \newpage
\section{Related Work}\label{rw}%
If any of the \NP-complete problems are shown to be in \P{} then all
of \NP-problems are in \P. In the same seminal paper
\cite{karp1972reducibility} where Karp determines a subset of the
\NP{} problems as \NP-complete with the aforementioned property, he
also gives twenty-one examples of such \NP-complete problems. Number
18 on the list is the knapsack problem: given a set
$M = \{m_1, m_2, \hdots, m_n\} \in \N^n, S \in \N$ find
$x \in \curl{0, 1}^n$ such that $Mx = S$. In other words, find a
subset of $M$ whose sum is equal to $S$.

The first cryptosystem to be based on an \NP-complete problem uses a
disguised knapsack problem and was attempted by Merkle and Hellman
\cite{merkle2019hiding}. We say a \emph{disguised} knapsack problem
since whether a cryptographic system can be as hard to break as to
solve an \NP-complete problem is an open problem in itself
\cite{pass2006parallel}. One might disguise an easy knapsack problem,
defined with a \emph{superincreasing}
${M_\text{easy}= \{m_k > 2m_{k-1} : m_k \in M\}}$ by modding integral
multiples of its elements. Note a solution $x$ for $M_\text{easy}$ can
be recovered easily by setting $x_k = 1$ if $S \geq m_k$ and then
subtracting $m_k$ from $S$. The same approach however does not work on
$M_\text{disguised} = \{Am_k \mod B : \gcd(A, B) = 1, m_k \in
M_\text{easy}\}$.

Lagarias and Odlyzko \cite{lagarias1985solving} showed that any
knapsack problem can be encoded as an \SVP. Take any knapsack problem
$(M, S)$ and the relevant solution $x$ such that $Mx = S$. Now
consider the lattice basis $\B_\text{bad}$ of dimension $d = n + 1$ in
Equation~\ref{knapsack}.  The lattice spanned by $\B_\text{bad}$ must
have a vector $t \in \L(\B_\text{bad})$ that is the result of an
integral linear combination due to $x$.
\begin{equation}\label{knapsack}
  t =
  \underbrace{
  \begin{bmatrix}
    2      & 0      & \cdots & 0      & 1      \\
    0      & 2      & \cdots & 0      & 1      \\
    \vdots & \vdots & \ddots & \vdots & \vdots \\
    0      & 0      & \cdots & 2      & 1      \\
    m_1    & m_2    & \cdots & m_n    & S      \\
  \end{bmatrix}
  }_{\B_\text{bad}}
  \col{x_1, x_2, \vdots, x_n, -1} =
  \col{2x_1-1, 2x_2-1, \vdots, 2x_n-1, M\cdot x-S} =
  \col{2x_1-1, 2x_2-1, \vdots, 2x_n-1, 0}
\end{equation}
Since $x$ is a binary vector, any $(2x_i-1) \in t$ must be either $1$
or $-1$. Therefore, $\norm{t} = \sqrt{n}$ and $t$ is very likely to be
the shortest vector in the lattice $\L(\B_\text{bad})$ because for all
other lattice vectors $v \in \L(\B_\text{bad})$ the length
$\norm{v} \gg \sqrt{n}$ due to the non-unit squares. The shortest
vector $t$ in the lattice spanned by $\B_\text{bad}$ will reveal the
solution $x$ to the knapsack problem \ie $x_k = 1$ if $t_k > 0$ else
$x_k = 0$.

\subsection{Lattice Reduction Algorithms}\label{sec:reduction}
If a lattice is expressed in terms of it's good basis then solving the
\SVP{} becomes fairly easy. This means that if we first solve the
\SBP{} then the \SVP{} is easy. For example, assume for a certain
$\B_\text{good}$ that all column vectors $b_i$ are pairwise orthogonal
\ie for $i \neq j$ we know that $b_i\cdot b_j = 0$. Then for any
$x \in \Z^d$,
$
  \norm{x_1b_1 + x_2b_2 + \cdots x_db_d}^2 =
  x_1^2\norm{b_1}^2 + x_2^2\norm{b_2}^2 + \cdots
  x_d^2\norm{b_d}^2
$
and the shortest non-zero vector(s) can be found in
$\curl{\pm b_1, \cdots \pm b_d}$. For an approximate solution of an
instance of the \CVP{} using a good basis, see Babai's nearest
hyperplane algorithm \cite{babai1986lovasz} or Theorem 7.34 (pg. 405)
of Hoffstein \cite{hoffstein2008introduction}.

\subsubsection{Lenstra, Lenstra, and Lov\'asz (\LLL) Algorithm}\label{sec:lll}
The first lattice reduction algorithm is by Gauss. It works like
Euclid's GCD algorithm but with two, two-dimensional vectors \ie
$\B = \{b_1, b_2\}$. Assume without the loss of generality that
$\norm{b_1} \leq \norm{b_2}$ then
$b_2 = b_2 - \near{(b_1 \cdot b_2)/\norm{b_1}^2}b_1$. If $\norm{b_2}$
is still greater than $\norm{b_1}$ we can stop. Otherwise, swap $b_2$
with $b_1$ and try again.

The \LLL{} algorithm generalises Gaussian lattice reduction from
two to $d$ dimensions. Just like Gaussian lattice reduction, it
subtracts an integral multiple of a shorter basis vector from a larger
basis vector until some size condition is fulfilled. For each
$1 \leq j < k \leq d$, we reduce $b_k$ as
$b_k = b_k - \near{\mu_{k,j}}b_j$ where
$ \mu_{k,j} = b_j^* \cdot b_k/b_j^* \cdot b_j^*.$ The vector $b_j^*$
is the $j^\text{th}$ basis vector in the Gram-Schmidt
orthogonalization of $\B$. The reduction of $b_k$ is carried out using
the Gram-Schmidt orthogonalizations of all the already reduced $b_j$
for every index $1 \leq j < k - 1$ where the \emph{size condition}
$\abs{\mu_{k,j}} > 0.5$ is met.

If \LLL{} terminates after only this recursive size reduction then the
goodness of the reduced basis depends on the order of the original
basis vectors in the basis matrix. Therefore, after the size reduction
of $b_k$, another condition, namely the popular \emph{Lov\'asz condition} is
checked;
for $\delta = 3/4$ we check the condition,
$ \norm{b_k^*}^2 \geq (\delta - \mu_{k,
    k-1}^2)\norm{b_{k-1}^*}^2.  $ If the Lov\'asz condition is met
then $b_k$ is considered reduced and $k$ will be
incremented. Otherwise, for the optimal ordering, we swap $b_k$ and
$b_{k-1}$ and decrement $k$\footnote{More precisely
  $k = \max(k - 1, 2)$}. \LLL{} is a polynomial time lattice reduction
algorithm for all $0 < \delta < 1$. The outer loop of Algorithm
\ref{lllalg} runs at most in
$\O(d^2\log(d) + d^2\log(\max\norm{b_i}))$ and solves the \appr\SVP{}
with an approximation factor of $\alpha = 2^{(d-1)/2}$. It is also an
open problem whether \LLL{} terminates in polynomial time for
$\delta = 1.$ For a more in-depth analysis, see Kalbach et
al. \cite{kalbach2024lll}.

\begin{algorithm}%
  \begin{flushleft}%
    \noindent\textbf{Input}: Lov\'asz condition constant:
    $0 < \delta < 1$ and the Bad basis: $\B = \{b_1, b_2, \dots, b_d\}$.
    
    \noindent\textbf{Output}: Good/Reduced basis: $\B = \{b_1, b_2, \dots, b_d\}$.
  \end{flushleft}%
  \begin{enumerate}[label=\arabic*:,noitemsep,partopsep=0pt,topsep=0pt,parsep=0pt]%
    \item $k \leftarrow 2$
    \item $(\B^*, \mu) \leftarrow \GM(\B)$
    \item $\quad \WHILE k \leq d$
    \item $\quad\quad \FOR j \in \{k - 1, k - 2, k - 3, \dots, 1\}$
    \item $\quad\quad\quad \IF \mu_{k, j} > 0.5$
    \item $\quad\quad\quad\quad b_k \leftarrow b_k - \near{\mu_{k, j}}b_j$
    \item $\quad\quad\quad\quad (\B^*, \mu) \leftarrow \GM(\B)$
    \item $\quad\quad \IF \norm{b_k^*}^2 \geq
            \paren{\delta - \mu_{k, k-1}^2}\norm{b_{k-1}^*}^2$
    \item $\quad\quad\quad k \leftarrow k + 1$
    \item $\quad\quad \ELSE$
    \item $\quad\quad\quad \SWAP(b_{k-1}, b_k)$
    \item $\quad\quad\quad (\B^*, \mu) \leftarrow \GM(\B)$
    \item $\quad\quad\quad k \leftarrow \max(k - 1, 2)$
  \end{enumerate}%
  \caption{%
    The \LLL{} algorithm (Section~\ref{sec:lll}). $\GM(\B)$ returns the
    Gram--Schmidt orthogonalisation of $\B$ as well as the projection
    scalars $\mu_{k, j}$ evaluated during the process of
    orthogonalising $b_k$ via $b_j$. $\SWAP(b_j, b_k)$ swaps
    $b_j, b_k$.%
  }\label{lllalg}%
\end{algorithm}%

\subsubsection{\LLL{} Variations}\label{sec:variation}
While \LLL{} is a polynomial time algorithm, many of it's
generalisations and extensions tend to perform just as fast during
empirical analysis and yield a further reduced basis
\cite{balny2025another}. Gama et al. argue by their extensive
empirical analysis \cite{gama2008predicting} that there is a gap
between what theory is able to prove and what is the true power of the
reduction algorithms.

The two possible exponential time variations of \LLL{} are DEEP (deep
insertion method) and BKZ (block Korkin--Zolotarev)
\cite{gama2008predicting}. DEEP differs from the standard \LLL{} when
the Lov\'asz condition fails.  Standard \LLL{} simply swaps the
$b_k$ with $b_{k-1}$ whereas in DEEP we insert $b_k$ at an optimal
place before the $k^\text{th}$ basis vector. While in BKZ, instead of
reducing $b_k$ with only one $b_{j}$, the same is done with a block of
basis vectors, $b_j, b_{j+1}, b_{j+2}, \dots b_{j+\beta - 1}$ where
$\beta$ is the block size. Note that if we let $\beta=d$ then the
shortest vector in the output of BKZ solves the \SVP{} problem and for
any $\beta < d$ we solve some version of the \appr\SVP{}.

\subsection{Sieving} Sieving for the shortest vector first surfaced
due to the works of Ajtai et al.  \cite{ajtai2001sieve}. At the time
of this writing, more than half ($437/847$) of the solutions posted at
the \SVP{} challenges website \cite{LatticeChallenge2025} are done via
sieving \cite{sun2020review}. The idea is: if we have $u, v \in \L$
and we want a shorter one, we might try $v - u$.  For a fixed
$P \subset \L$, repeatedly check if there exists a pair $u, v \in P$
such that $\norm{v - u} < \norm{v}$ then $v$ is replaced with $v - u$.
Laarhoven \cite{laarhoven2019} set sieving up as a simple genetic
algorithm. The foremost concern here arises about a possible method of
mutation if lattice vectors $v \in \L(\B)$ are treated as individuals
in a genetic algorithm. This is due to the fact that mutating a gene
$v_i \in v$ may cause $v$ to step outside of the lattice. For this
reason, they encode the individual for $v = \B x$ as $x \in
\Z^d$. This is possible because of the definition of a lattice given
in Section \ref{prelim}. Now $x$ maybe perturbed by adding binary
noise to a random $x_i \in x$. While encoding a lattice vector
$v = \B x$ as $x$ gives a way to mutate $v$ through $x$, it requires a
matrix multiplication $\B x$ in order for the fitness (norm) of $v$ to
be calculated.


\section{Methodology}\label{sec:method}

\begin{figure}[hbtp]%
  \centering%
  \subfloat[Iteration 1.]{\includegraphics[width=0.48\linewidth]{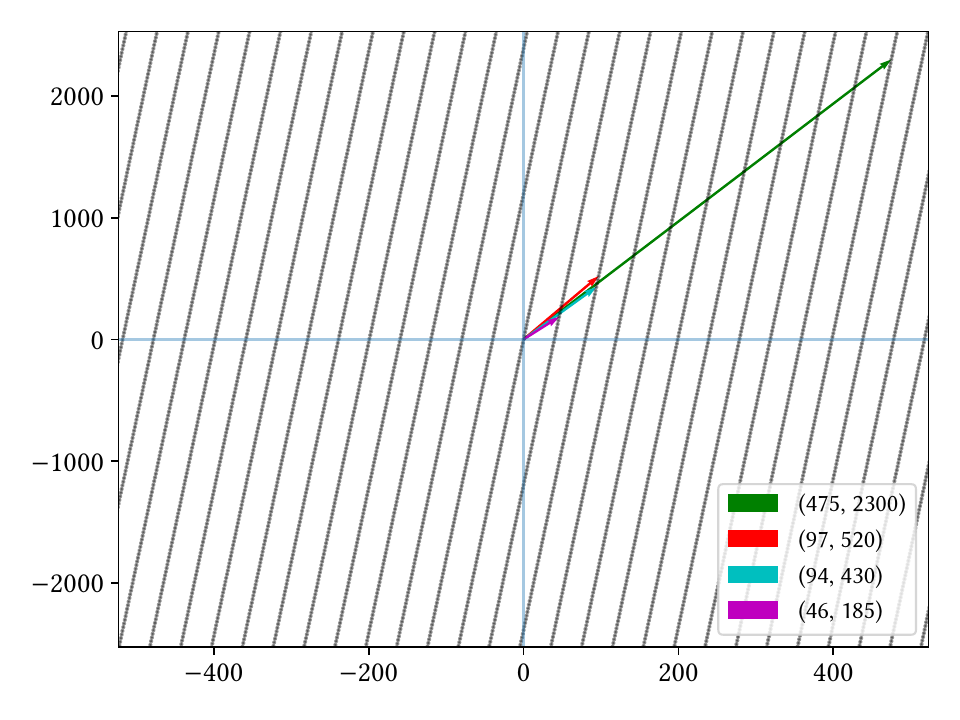}}
  \subfloat[Iteration 2.]{\includegraphics[width=0.48\linewidth]{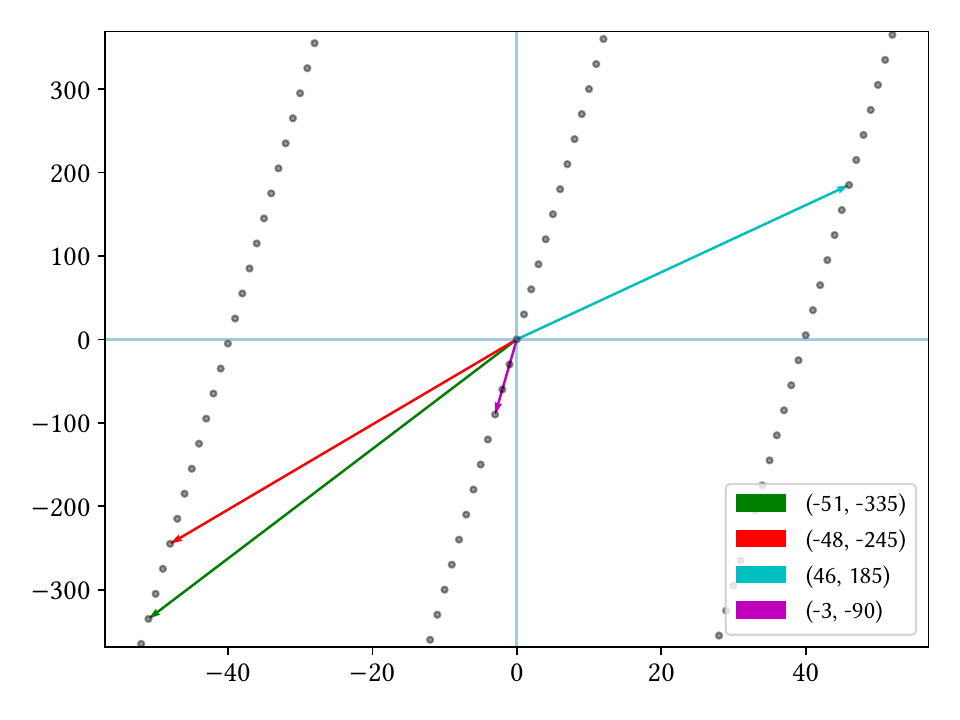}}\\
  \subfloat[Iteration 6.]{\includegraphics[width=0.48\linewidth]{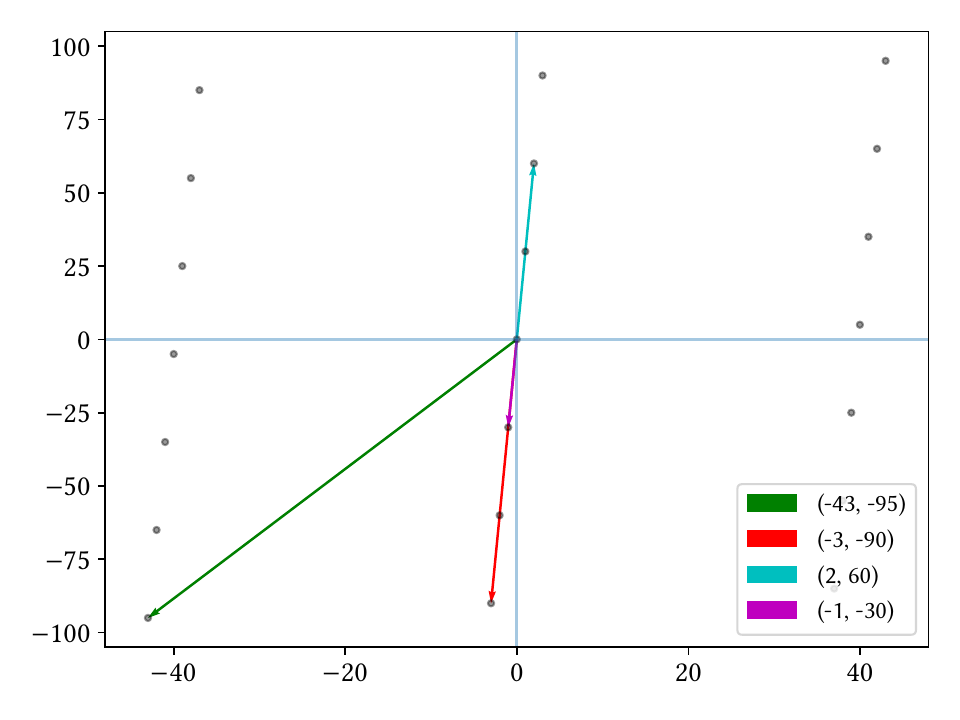}}
  \subfloat[Iteration 7.]{\includegraphics[width=0.48\linewidth]{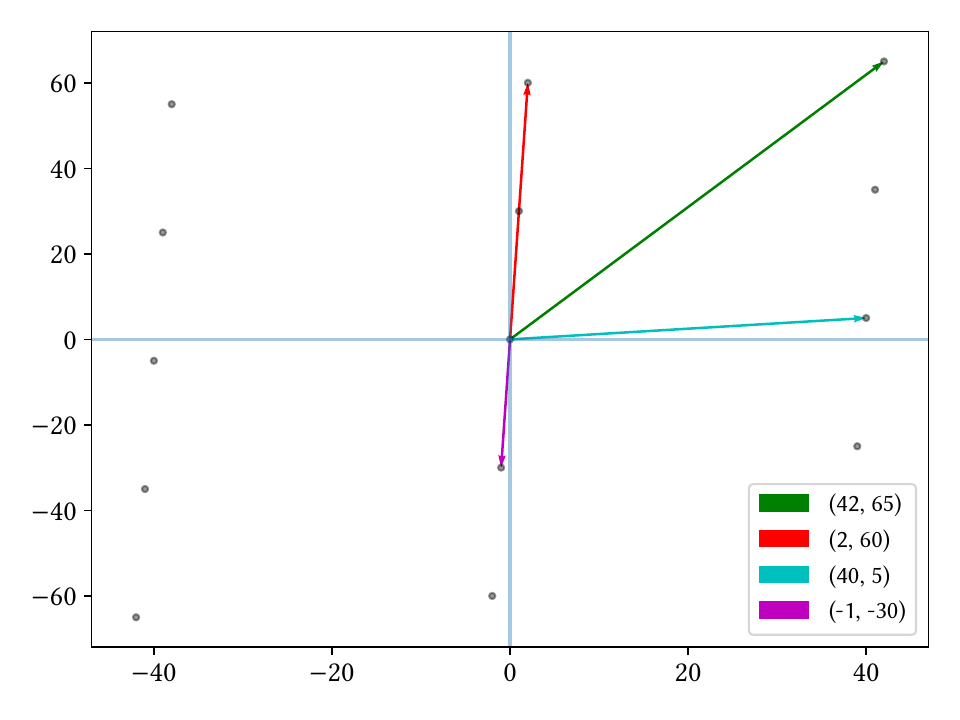}}
  \caption{%
    Algorithm~\ref{naive} reducing $\B_\text{bad}$ in
    Eq.~\ref{running} starting with population $P$ in
    Eq.~\ref{population}. Note that in this case, we find the shortest
    vector by solving the \SBP{} exactly. The seventh iteration in
    Figure~\ref{naivedemo} contains $\B_\text{good}$ of
    Equation~\ref{running}.
  }%
  \label{naivedemo}%
\end{figure}

We give a na\"ive sieving Algorithm~\ref{naive}. It checks the
difference of all the possible pairings in the current population of
vectors $P$ for a shorter non-zero vector not already present in
$P$. These newer and shorter vectors are collected in $R$. At the
start of each iteration, we combine $P$ and $R$, \emph{selecting} out
of the combination at most $|P|$ successive shortest vectors to be
reassigned as $P$. The sieving terminates when $P^2$ no longer
contains a pair $(u, v)$ whose difference's length is shorter than any
of the ones already in $P$.  We show an example by reducing the bad
basis $\B_\text{bad}$ given in eq.~\ref{running} of Sec.~\ref{prelim}.
The initial $P$ is given as,
\begin{equation}\label{population}
  P =
  \begin{bmatrix}
    46  & 94  & 97  & 475 \\
    185 & 430 & 520 & 2300
  \end{bmatrix}
\end{equation}
and Figure~\ref{naivedemo} shows the new $P$ at select iterations of a
total of seven iterations that were ran.

\begin{algorithm}
\begin{flushleft}
    \noindent\textbf{Input}: $P \subset \L$. \\
    \noindent\textbf{Output}: Reduced version of subset: $P.$
  \end{flushleft}
  \begin{enumerate}[label=\arabic*:,noitemsep,partopsep=0pt,topsep=0pt,parsep=0pt]%
    \item $\DO$
    \item $\quad R \leftarrow \nil$
    \item $\quad \FOR \text{each } (u, v) \in P^2$
    \item $\quad\quad t \leftarrow v - u$
    \item $\quad\quad \IF (\vec{0} \neq t \not\in P) \wedge
    (\norm{t} < \norm{u} \vee \norm{t} < \norm{v})\;\THEN R \leftarrow R \cup \{t\}$
    \item $\quad P \leftarrow \SL(P \cup R)$
    \item $\WHILE R \neq \nil$
  \end{enumerate}%
  \caption{%
    Na\"ive sieving algorithm with global selection.
    $\SL(P \cup R)$ selects $|P|$ shortest (fittest) vectors
    (individuals) among $P \cup R$.
  }
  \label{naive}
\end{algorithm}

\begin{figure}[hbtp]%
  \centering%
  \subfloat[%
    Illustration of our crossover technique vs. Laarhoven's. See Section
   ~\ref{sec:cross}.%
  ]{\includegraphics[width=0.48\linewidth]{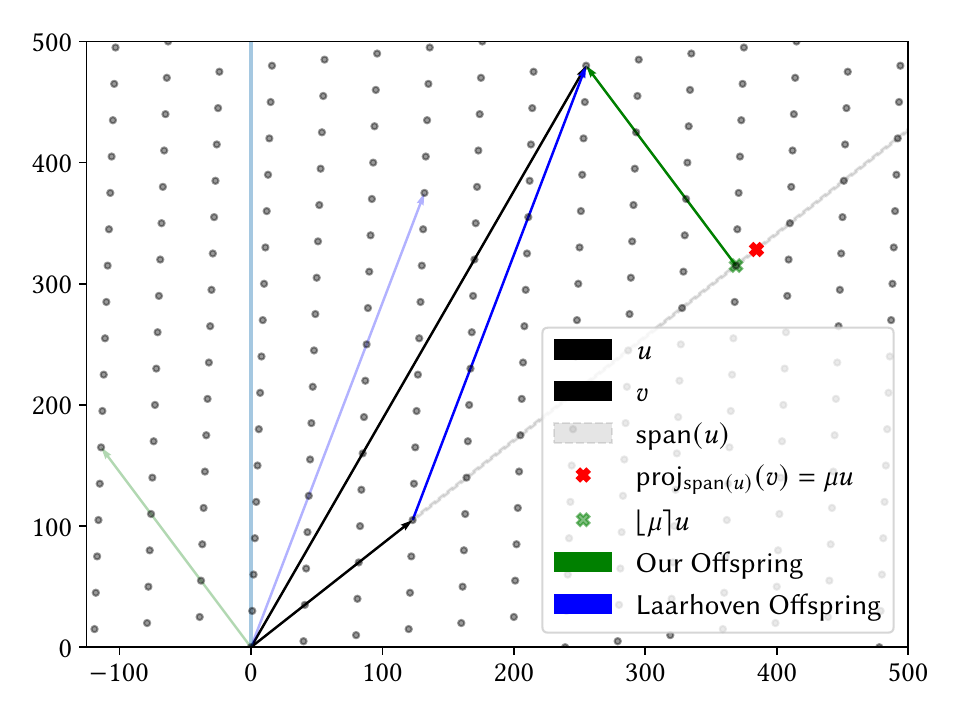}\label{fig:cross}}%
  \hfill%
  \subfloat[%
    Fig.~\ref{fig:cross} offspring mutations. See Section~\ref{sec:mutation}.
  ]{\includegraphics[width=0.48\linewidth]{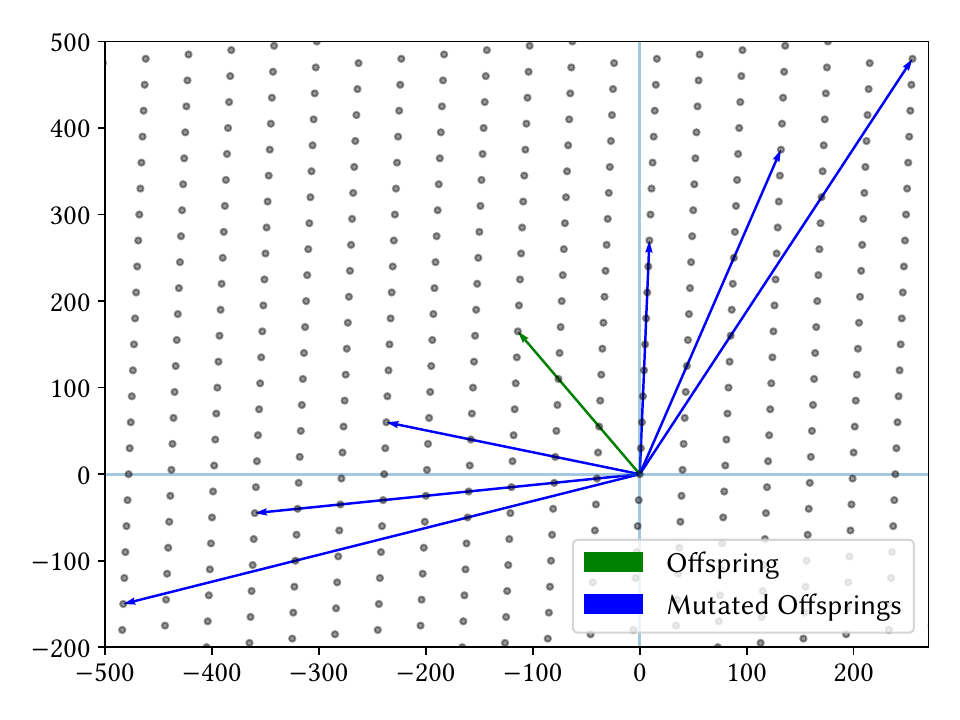}\label{fig:mutate}}%
  \caption{%
    Illustration of crossover and mutation techniques from Section~\ref{sec:ga}.%
  }%
  \label{fig:cross-mutate}%
\end{figure}%

\subsection{Our Genetic Algorithm for Sieving}\label{sec:ga} Following is a
specification for each of the components of the genetic
algorithm~\ref{gas} used in this paper to solve the \appr\SVP.

\paragraph{Genotype} We represent vectors
$v = \arr{v_1, v_2, \dots, v_d} \in \L$ as is. Note that here
$\L \subseteq \Z^{d}$ or $\L \subseteq \G^{d}$ (see
Section~\ref{sec:module}).

\paragraph{Fitness Function} We evaluate the fitness of a vector $v$
by the inverse of its $\ell^2$ norm: $\FIT(v) =
\norm{v}_2^{-1}$. Observe that one may also invert the squared norm.

\paragraph{Crossover}\label{sec:cross}
For parents, $u, v \in P$, we may generate the offspring $t$,
\[
  t = \CROSS(u, v) = v - \near{\mu} u
  \quad
  \text{where}
  \quad
  \mu = \frac{\Re(u\cdot v)}{u\cdot u} + \frac{\Im(u\cdot v)}{u\cdot u}i.
\]
If $u, v \in \Z^d$ then the imaginary part $\Im(u\cdot v) = 0$. We
round off a complex number $z = \alpha + \beta i$ where
$(\alpha, \beta) \in \R^2$ by
$\near{z} = \near{\alpha} + \near{\beta}i$. See Figure~\ref{fig:cross}
for a geometric interpretation of the crossover in 2 dimensions. The
idea is that if $(u, v)$ are near to each other then the offspring $t$
is of high fitness but in the average case, $(u, v)$ are not
near. Letting $u$ be the shorter vector we realise that multiple
copies $\near{\mu}$ of $u$ could be subtracted from $v$. This constant
$\mu$ is used to evaluate the projection of $v$ onto the span of $u$
\ie ${\proj_{\spn(u)}(v) = \mu u}$. However, as seen in Figure
\ref{fig:cross}, $\proj_{\spn(u)}(v)$ is not necessarily in the
lattice. Therefore, instead of $\proj_{\spn(u)}(v)$ we subtract
$\near{\mu} u$ from $v$ getting a much shorter (fitter) offspring than
the Laarhoven's offspring $v - u$.

\paragraph{Mutation}\label{sec:mutation} The $j^\text{th}$ column of the initial
population and the subsequent generations is given as $P_j = \B c_j$.
Laarhoven \cite{laarhoven2019} represent each individual in the
population as $c_j$ instead of $\B c_j$. This enabled them to mutate
$P_j = \B c_j$ by adding a small integral perturbation to
$c_{i,j}$. However, this representation requires a matrix
multiplication before the fitness (inverse $\ell^2$ norm) of each
individual can be calculated. In our representation, if an offspring
was to be mutated, we may multiply $\mu$ (Section~\ref{sec:cross}) by
a normal random variable $\xi \sim \mathcal{N}(1, 1)$ of mean and
standard deviation $1$, i.e.,
$t_\text{mutated} = v - \near{\xi\mu} u$.  The crossover step can now
be given as $\CROSS(u, v) = v - \near{\xi\mu} u$ where either
$\xi = 1$ or $\xi \sim \mathcal{N}(1, 1)$ under a small
probability. As an example, for each
$\xi \in \{0.1, 0.4, 0.7, 1, 1.3, 1.6, 1.9\}$, Figure~\ref{fig:mutate}
shows the corresponding six possible mutations of our offspring shown
in Figure~\ref{fig:cross}.

\paragraph{Initial Population} Before we proceed with the generation
of the initial population using $\B$, we may \emph{optionally} compute
the \emph{Hermite normal form} of $\B$ as well as reduce it using the
\LLL{} algorithm for some $\delta$. The Hermite normal form is to
integral matrices what the reduced echelon form is for the matrices
over the reals. Let $n = |P|$, we generate the initial population via
a constant $d$ by $n$ binary matrix $C$ like this: ${P = \B C}$. If
$\L(\B) \subseteq \Z^{d}$ or $\B \in \Z^{d\times d}$, then let
$C \in \{0, 1\}^{d\times n}$ where $\text{Pr}(C_{i,j} = 1) = \rho.$
However, if $\L(\B) \subseteq \G^{d}$ or $\B \in \G^{d\times d}$ then
for $\alpha, \beta \in \{0, 1\}$ we have $C_{i,j} = \alpha + \beta i$
where $ \text{Pr}(\alpha = 1) = \text{Pr}(\beta = 1) = \rho$.

\paragraph{Selection Strategy} Let $P$ be the previous and $R$ be the
current generation. The next generation is then produced by picking
the $|P|$ successive shortest vectors from the pool $P \cup R$ and
then assigning them back to ${P \leftarrow \Elite(P \cup R)}$. $P$
will now be sorted by vector length in ascending order. We will use
this $P$ to select the individuals for crossover. Let $u_i, v_j$ be
the $i^\text{th}$ and $j^\text{th}$ vectors in $P$ then
Algorithm~\ref{sl} states how we can select pairings of vectors in
$P$.

\begin{algorithm}
  \begin{flushleft}
    \noindent\textbf{Input}: Population: $P$. \\
    \noindent\textbf{Output}: Generated pairs for reproduction.
  \end{flushleft}
  \begin{enumerate}[label=\arabic*:,noitemsep,partopsep=0pt,topsep=0pt,parsep=0pt]%
    \item $\FOR i \in \{1, 2, 3, \dots |P| - 1\}$
    \item $\quad \FOR j \in \{i + 1, \dots |P|\}$
    \item $\quad\quad \YIELD (u_i, v_j)$
  \end{enumerate}%
  \caption{%
    ${\YIELD (u_i, v_j)}$ yields each pair to $\SL(P)$ in Algorithm~\ref{gas}.%
  }
  \label{sl}
\end{algorithm}

\begin{algorithm}[hbtp]%
  \begin{flushleft}
    \noindent\textbf{Input}:
    (1) Basis: $\B$, (2) Population size: $n$ (3) Sampling density:
    $\rho$
    
    \noindent\textbf{Output}: Approximation of the shortest vector.
  \end{flushleft}
  \begin{enumerate}[label=\arabic*:,noitemsep,partopsep=0pt,topsep=0pt,parsep=0pt]%
    \item $C_{d,n} \leftarrow (c_{i,j} \sim \text{Bernoulli}(\rho))$
    \item $P \leftarrow \B C$
    \item $R \leftarrow \nil$
    \item $\DO$
    \item $\quad P \leftarrow \Elite(P \cup R)$
    \item $\quad R \leftarrow \nil$
    \item $\quad \DO$
    \item $\quad\quad u, v \leftarrow \SL(P)$
    \item $\quad\quad t \leftarrow \CROSS(u, v)$
    \item $\quad\quad \IF (\vec{0} \neq t \not\in P) \wedge
    (\norm{t} < \norm{u} \vee \norm{t} < \norm{v})$
    \item $\quad\quad\quad R \leftarrow R \cup \{t\}$
    \item $\quad \WHILE |R| < n^{1.5}$
    \item $\WHILE \forall v \in P, \FIT(v) < w^{-1}$
  \end{enumerate}
  \caption{%
    Our genetic sieving algorithm. See Section~\ref{sec:ga}.%
  }%
  \label{gas}
\end{algorithm}

\subsection{Analysis of the GAs for Sieving}\label{sec:analysis}

Let $m$ count the max. generations, $n = \abs{P}$ be the size of the
initial population and $d$ be the dimension of the lattice. On
average, our algorithm runs in $\O(dmn^{1.5})$. This due to the
$\O(d)$ cost of each fitness evaluation times the total number of
generations $m$ times the total number of expected $n^{1.5}$
crossovers.

In the case of Laarhoven's \cite{laarhoven2019} algorithm, we keep the
same $m$ generations, $n$ individuals and note that they consider all
possible children for a potential insertion in the next
generation. Furthermore, they represent the $j^\text{th}$ individual
$P_j = \B c_j$ as $c_j$ in contrast to $\B c_j$ to stay compatible
with their mutation operation. Their representation of $P_j$ as $c_j$
introduces a matrix multiplication (namely $\B$ multiplied on the left
with $ c_j$) each time fitness is to be calculated since fitness is a
function of $\norm{\B c_j}$. Therefore, Laarhoven's algorithm can be
shown to run in $\O(d^3mn^2)$ on average. The exponent $1.5$ of
$n^{1.5}$ in our algorithm was determined by a combination of
empirical analysis and heuristics from the original choices of
$\delta$ in the \LLL{} (Section~\ref{sec:lll}).


\section{Results}\label{results}%
Recall that lattice-based cryptography relies on the inability of
lattice reduction algorithms to solve the \appr\SVP{} in high
dimensions $(d > 300)$ with an approximation factor
$\alpha = \O(\sqrt{d})$. For dimensions $d \leq 100$ in integral and
$d \leq 50$ in module lattices, we optimally solve the \appr\SVP{}
with an $\alpha < 2.05$ (Table~\ref{tab:result}).

\sloppy Detailed results for reduction on \SVP{} challenges
\cite{LatticeChallenge2025} in dimensions $40 \leq d \leq 100$ are
given in Section~\ref{svpc}. Parallelly, we generated random integral
lattices of equal dimensions to further test the performance of
Algorithm \ref{gas}. Results on these lattices are given in Section
\ref{rnd}. Finally, we reduce module lattices of the dimensions 20,
30, 40 and 50, giving the results in Section~\ref{ml}.

The experiments reducing the \SVP{} challenge and the random integral
lattices in dimensions 40, 50, 60, 70 and 80 were carried out
twice. Once with no mutations and once with $1\%$ chance of
mutation. The resulting shortest vectors were the same in either case.
Although, the experiments with mutations took about twice as
long. Afterwards, experiments on lattices of dimensions
$d \in \{90, 100\}$ and module lattices were therefore carried out
without mutations.

\begin{figure}[hbtp]%
  \captionsetup[subfigure]{justification=centering}%
  \centering%
  \subfloat[%
    Shortest vector length comparison in the \SVP{} challenge lattices.
  ]{\includegraphics[width=0.48\linewidth]{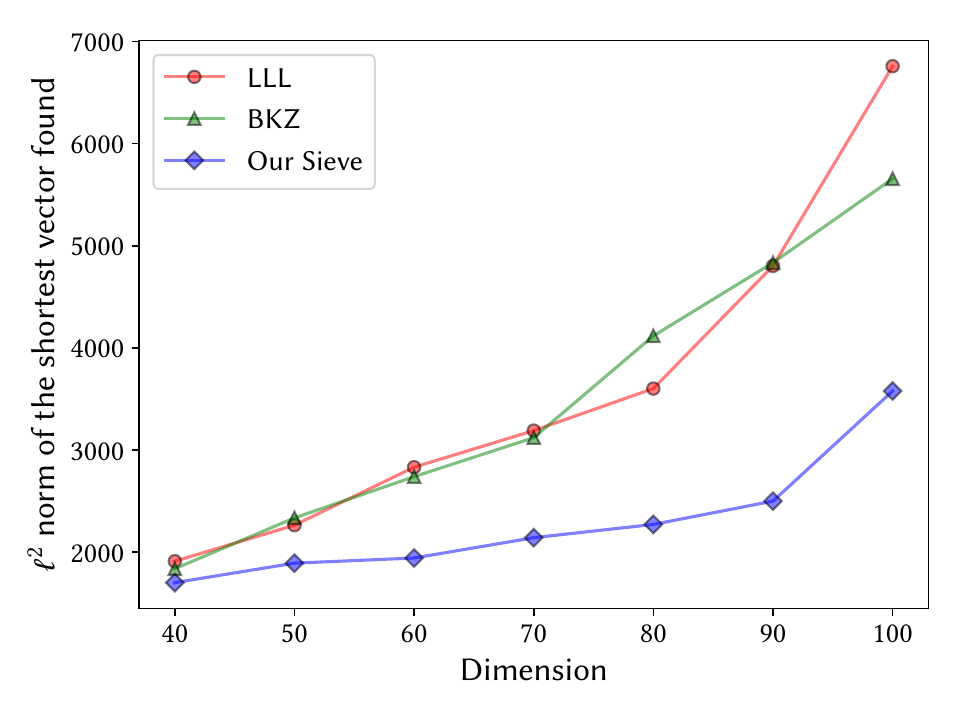}}%
  \hfill
  \subfloat[%
    Shortest vector length comparison in the randomly generated lattices.
  ]{\includegraphics[width=0.48\linewidth]{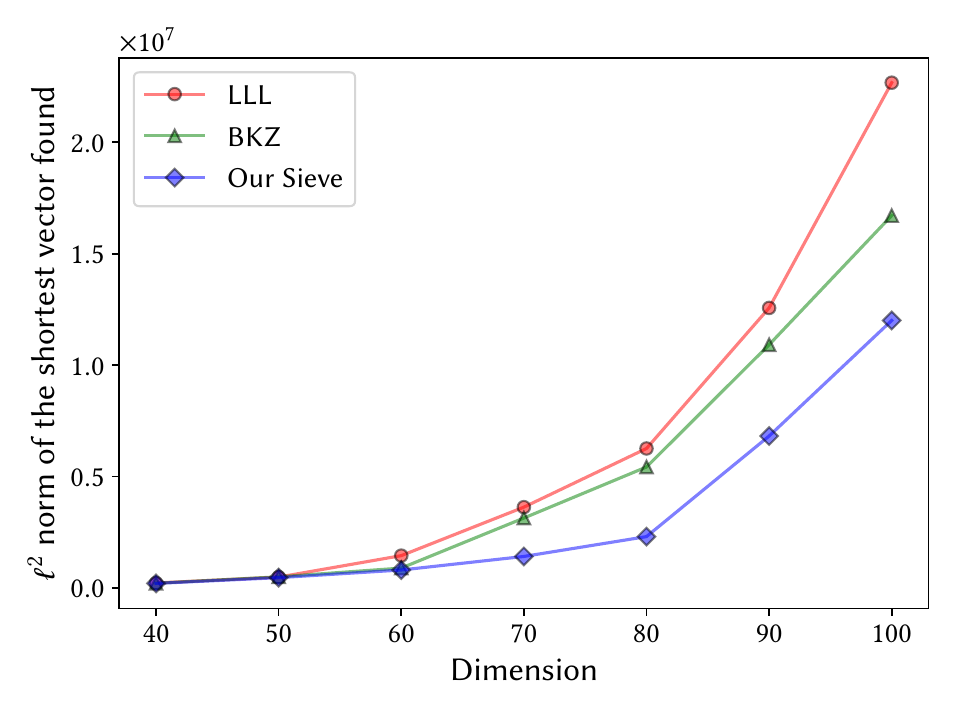}}%
  \caption{%
    Comparison of the shortest vector length found by different algorithms.
  }%
  \label{img:compare}%
\end{figure}%

The stopping condition for each row in the results of Table
\ref{tab:result} can be shown by letting $w$ of Algorithm~\ref{gas} be
$\ell$ (the shortest/most fit vector found). In practice, we stopped
once the difference of the mean population length between the current
and the last generation was consecutively less than 1 for three
generations.

\subsection{Reducing \SVP{} Challenge Lattices}\label{svpc}
We started by reducing each of the lattices using the \LLL{} algorithm
as described in Section~\ref{sec:lll} for $\delta = 1-10^{-7}$. Note
that the initial population included this reduced $\B$ as is. This
implies that the algorithm must find something better than the final
results of the $\LLL{}$ for a $\delta=0.9999999$ fairly close to 1.

For the Algorithm~\ref{gas} parameter $\rho = 0.01$, Table
\ref{tab:result} summarises the results at the final step for each
$40 \leq d \leq 100$. Up until 80 dimensions, our results (length of
the shortest vector) match the best ones found so far. Furthermore,
these were all solved on an M2 Macbook Air with 24 GB of memory.

\subsection{Reducing Random Integral Lattices}\label{rnd} We
generate these lattices $\L(\B)$ by uniformly randomly generating each
element $b_{i,j}$ of $\B$ between $-d^3$ and $d^3$.  Afterwards, we
compute the Hermite normal form (this is to integral matrices what the
reduced echelon form is for the matrices over the reals) of $\B$ and
reduce it with the \LLL{} algorithm for $\delta = 1-10^{-7}$.

For the Algorithm~\ref{gas} parameter $\rho = 0.01$, Table
\ref{tab:result} summarises the results at the final step for each
$40 \leq d \leq 100$. We optimally solve the \appr\SVP{} for
$\alpha < 1.5$ for all $40 \leq d \leq 100$.

\subsection{Reducing Module Lattices}\label{ml}
We generate these lattices $\L(\B)$ by generating each
$\alpha + \beta i \in \B$ containing $\alpha$ and $\beta$ picked
uniformly randomly between $-d^3$ and $d^3$. Furthermore, to each
basis vector $b_i \in \B$ we add $d/4$ many other randomly chosen
basis vectors $b_j \in \B$.

For the same as before Alg.~\ref{gas} parameter $\rho = 0.01$, Table
\ref{tab:result} summarises the results at the final step. We solve
the \appr\SVP{} for $\alpha < 2.05$ for all $20 \leq d \leq 50$.

\subsection{Results Summary \& Discussion}\label{disc}
Among the \SVP{} Challenge Lattices, for $d \leq 80$ dimensions, our
resultant short vectors match the known best ones found so far. For
all of the integral and module lattices, we optimally solve the
\appr\SVP{} with $\alpha < 1.5$, and $\alpha < 2.05$.

We preprocessed the basis with the state of the art $\LLL{}$ algorithm
for a $\delta=0.9999999$ fairly close to 1. This puts our algorithm in
direct comparison with the $\LLL{}$ reduction algorithm. Figure
\ref{img:compare} further compares the reduction power of $\LLL{}$ as
well as BKZ with $\delta=0.9999999$ and the standard block-size of 10
against our algorithm. We outperform both the $\LLL{}$ and the BKZ.

Laarhoven solved the same (and only attempted) 40 dimensional lattice
as we did in our results. They reported 29 iterations starting with a
set of 1500 lattice vectors to find the shortest vector of norm 1702
and an average vector of norm 2008 \cite{laarhoven2019}. Our approach
took 7 iterations starting with a set of 1000 lattice vectors to find
the shortest vector of norm 1702 and an average vector of norm
1981. Laarhoven found the same shortest vector as we did in 1.9
seconds compared to the 0.06 seconds it took our approach. While we
can not faithfully explain the difference in time due to the possible
machine/implementation differences, we suspect the difference is due
the absence of the matrix multiplication overhead in our algorithm as
explained in Sections~\ref{sec:mutation} and~\ref{sec:analysis}. The
reduction in the number of generations can be explained by the
optimisations inspired by the classical algorithms in our crossover
operator (Section~\ref{sec:cross}).

Evolutionary sieving for the shortest vector remains a relatively new
and under-explored area. While we presently do not anticipate it
posing a direct threat to existing cryptographic algorithms, this
paper demonstrates that interdisciplinary approaches, combining
machine learning and number theory, can effectively reduce a lattice
of 100 dimensions with a modest approximation factor. This promising
intersection of fields opens up exciting future avenues.


\section{Conclusion and Future Work}\label{sec:conclusion}%
Sieving algorithms have become one of the most practical methods for
the \SVP{}.  We improved Laarhoven's \cite{laarhoven2019}
representation of the \SVP{} for GAs by reducing the number of
generations and population size required to obtain the same (or better
average) results. Our improvements are due to the refined genotype and
crossover born out of the insights from the classic number theoretic
algorithms. We further showed the proposed algorithm's reduction
potential on the increasingly significant module lattices.

It is important to recognize that scaling \emph{any} method for
dimensions beyond 100 presents considerable challenges due to the
inherent complexities of lattice problems. This difficulty underscores
the reliance of the state-of-the-art cryptographic schemes on such
problems
\cite{alagic2019status,alagic2020status,alagic2022status,alagic2025status}.

We can advance our understanding of the reduction capabilities of
evolutionary approaches in dimensions greater than 100 in various
ways. For instance, managing the population as described by
\cite{goldman2014parameter,bosman2016expanding} or selection through
techniques such as tournament selection could yield better
results. Additionally, implementing more sophisticated crossover
strategies, such as performing multiple steps of Gauss reduction on
parent vectors, or employing a nearest neighbour search for parent
selection could enhance the effectiveness of these approaches.

Furthermore, evolutionary sieving techniques could be investigated
within module lattices in rings beyond Gaussian integers, such as more
general Cyclotomic fields. This multifaceted exploration holds the
potential to significantly improve our ability to address the \SVP{}
in higher dimensions, thereby reinforcing the robustness of
evolutionary sieving in integral and module lattices.


\setlength{\tabcolsep}{3pt}
\begin{table}[hbtp]%
  \caption{%
    A comparison of the shortest vector length in each dimension ($d$)
    found via the Algorithm~\ref{gas} ($\ell$) against the expected
    shortest length due to the Gaussian Heuristic.  Recall from
    Section~\ref{sec:gh} that the exact length of the shortest
    non-zero vector is unknown in the general case and therefore, for
    a sufficiently large $d$ \cite{hoffstein2008introduction}, we may
    rely on the Gaussian expected shortest length ($\sigma$).  By
    Definition~\ref{def:alpha}, $\ell/\sigma \geq \alpha$. The column
    $g$ is the total number of generations evolved and $n$ gives the
    size of the initial population.%
  }
  \label{tab:result}\vspace{0.25\baselineskip}
  \begin{tabular}{@{}r|rrrrr|rrrrr@{}}
    \toprule
        & \multicolumn{5}{c}{\SVP{} \sc{challenge lattices}} & \multicolumn{5}{|c}{\SVP{} \sc{random integral lattices}}                                                                                            \\
    \midrule
    $d$ & $\sigma$                                           & Alg.~\ref{gas} $(\ell)$                                   & $\alpha$ & $g$ & $n$     & $\sigma$ & Alg.~\ref{gas} $(\ell)$ & $\alpha$ & $g$ & $n$     \\
    \midrule
    40  & 1560                                               & 1702.46                                                   & 1.09     & 7   & 1000    & 212828   & 201343.89               & 0.95     & 10  & 800     \\
    50  & 1746                                               & 1893.17                                                   & 1.08     & 8   & 1250    & 522364   & 460101.23               & 0.88     & 13  & 1000    \\
    60  & 1916                                               & 1943.40                                                   & 1.01     & 16  & 9900    & 1113319  & 803170.35               & 0.72     & 25  & 3000    \\
    70  & 2065                                               & 2142.60                                                   & 1.04     & 11  & 49980   & 2071820  & 1411636.20              & 0.68     & 18  & 28000   \\
    80  & 2205                                               & 2272.19                                                   & 1.03     & 16  & 240000  & 3351770  & 2302176.05              & 0.69     & 16  & 160000  \\
    90  & 2544                                               & 2500.40                                                   & 0.98     & 25  & 1638000 & 5564035  & 6815686.92              & 1.22     & 18  & 540000  \\
    100 & 2468                                               & 3578.23                                                   & 1.45     & 16  & 1500000 & 8339504  & 12002723.15             & 1.44     & 16  & 1000000 \\
    \bottomrule
  \end{tabular}%
  \centering\setlength{\tabcolsep}{6pt}
  \begin{tabular}{r|rrrrr}
        & \multicolumn{5}{c}{\sc{module lattices}}                                                   \\
    \midrule
    $d$ & $\sigma$                                 & Alg.~\ref{gas} $(\ell)$ & $\alpha$ & $g$ & $n$  \\
    \midrule
    20  & 22947                                    & 26188.63                & 1.14     & 5   & 2000 \\
    30  & 136309                                   & 177814.21               & 1.30     & 4   & 3000 \\
    40  & 369551                                   & 754691.69               & 2.04     & 4   & 4000 \\
    50  & 863938                                   & 1482020.58              & 1.72     & 5   & 5000 \\
    \bottomrule
  \end{tabular}
\end{table}%


\begin{credits}
  \subsubsection{\ackname} We thank Dimitrios Diochnos and Dean Hougen
  for their constructive criticism and other valuable feedback on the
  manuscript and acknowledge the U. S. National Science Foundation for
  their support through the grant \texttt{CCF-2530361}.
  
  \subsubsection{\discintname} The authors have no competing interests
  to declare that are relevant to the content of this article.
\end{credits}


\bibliographystyle{splncs04}
\bibliography{tex/citation}
\end{document}